\documentclass[12pt]{article}
\usepackage{amsmath}
\usepackage{graphicx}
\usepackage{bm}
\usepackage{fancyhdr}
\usepackage{amssymb}
\begin{document}



\def\a{\alpha}
\def\b{\beta}
\def\d{\delta}
\def\e{\epsilon}
\def\g{\gamma}
\def\h{\mathfrak{h}}
\def\k{\kappa}
\def\l{\lambda}
\def\o{\omega}
\def\p{\wp}
\def\r{\rho}
\def\t{\tau}
\def\s{\sigma}
\def\z{\zeta}
\def\x{\xi}
\def\V={{{\bf\rm{V}}}}
 \def\A{{\cal{A}}}
 \def\B{{\cal{B}}}
 \def\C{{\cal{C}}}
 \def\D{{\cal{D}}}
\def\K{{\cal{K}}}
\def\O{\Omega}
\def\R{\bar{R}}
\def\T{{\cal{T}}}
\def\L{\Lambda}
\def\f{E_{\tau,\eta}(sl_2)}
\def\E{E_{\tau,\eta}(sl_n)}
\def\Zb{\mathbb{Z}}
\def\Cb{\mathbb{C}}

\def\R{\overline{R}}

\def\beq{\begin{equation}}
\def\eeq{\end{equation}}
\def\bea{\begin{eqnarray}}
\def\eea{\end{eqnarray}}
\def\ba{\begin{array}}
\def\ea{\end{array}}
\def\no{\nonumber}
\def\le{\langle}
\def\re{\rangle}
\def\lt{\left}
\def\rt{\right}

\newtheorem{Theorem}{Theorem}
\newtheorem{Definition}{Definition}
\newtheorem{Proposition}{Proposition}
\newtheorem{Lemma}{Lemma}
\newtheorem{Corollary}{Corollary}
\newcommand{\proof}[1]{{\bf Proof. }
        #1\begin{flushright}$\Box$\end{flushright}}

\baselineskip=20pt

\newfont{\elevenmib}{cmmib10 scaled\magstep1}
\newcommand{\preprint}{
   \begin{flushleft}
   \end{flushleft}\vspace{-1.3cm}
   \begin{flushright}\normalsize
   \end{flushright}}
\newcommand{\Title}[1]{{\baselineskip=26pt
   \begin{center} \Large \bf #1 \\ \ \\ \end{center}}}
\newcommand{\Author}{\begin{center}
   \large \bf
Xin Zhang${}^{a}$,~Junpeng Cao${}^{a, b}$,~Wen-Li Yang${}^{c, d}$,~Kangjie
Shi${}^c$ ~and~Yupeng Wang${}^{a,b} \,\footnote{Corresponding author:
yupeng@iphy.ac.cn}$
 \end{center}}
\newcommand{\Address}{\begin{center}
     ${}^a$Beijing National Laboratory for Condensed Matter
           Physics, Institute of Physics, Chinese Academy of Sciences, Beijing
           100190, China\\
     ${}^b$Collaborative Innovation Center of Quantum Matter, Beijing,
     China\\
     ${}^c$Institute of Modern Physics, Northwest University,
     Xian 710069, China\\
     ${}^d$Beijing Center for Mathematics and Information Interdisciplinary Sciences, Beijing, 100048,  China
   \end{center}}
\newcommand{\Accepted}[1]{\begin{center}
   {\large \sf #1}\\ \vspace{1mm}{\small \sf Accepted for Publication}
   \end{center}}

\preprint \thispagestyle{empty}
\bigskip\bigskip\bigskip

\Title{Exact solution of the one-dimensional super-symmetric t-J
model with unparallel boundary fields} \Author

\Address \vspace{1cm}

\begin{abstract}
The exact solution of the one-dimensional super-symmetric t-J model
under generic integrable boundary conditions is obtained via the
Bethe ansatz methods. With the coordinate Bethe ansatz, the
corresponding $R-$matrix and $K-$matrices are derived for the second
eigenvalue problem associated with spin degrees of freedom. It is
found that the second eigenvalue problem can be transformed into that
of the transfer matrix of the inhomogeneous XXX spin chain, which
allows us to obtain  the spectrum of the Hamiltonian and the associated
Bethe ansatz equations by  the off-diagonal Bethe ansatz
method.

\vspace{1truecm} \noindent {\it PACS: 71.10.Fd; 75.10.Lp; 71.10.Pm}


\end{abstract}

\newpage

\section{Introduction}

The $t$-$J$ model is one of the most important models for describing
the strongly correlated electronic systems and especially for the
high-$T_c$ superconductivity \cite{zhang,Anderson2,Essler2}. The model is in fact a
large $U$ limit of the well-known Hubbard
model \cite{Hu,Eskes,McMahan,Hybertsen1,Hybertsen2} and provides a
non-phonon mechanism for the high-$T_c$
superconductivity \cite{Anderson,Ogata,Putikka}. Interestingly, the
model in one spatial dimension at the supersymmetric points
$2t=\pm{J}$ \cite{Wiegmann,Foerster,Foe92,Foerster2} is exactly solvable
\cite{Sutherland,Schlotmann,Schulz,Essler}. Based on this observation, some
important physical properties such as the elementary excitations \cite{Bares,Mishchenko,Martinez},
correlation functions \cite{Kawakami2} and the thermodynamics \cite{Williams,Juttner,Giamarchi,Emery} were
studied by many authors. The model with diagonal boundary fields in the spin sector (or in the charge sector with boundary
chemical potentials) has been studied extensively by the nested Bethe ansatz method
\cite{Foerster2,Gon94,Ess96,wang,Fan,Zhou,Fan2,Fan3,Bedurftig,Hu2,Gal07} or the off-shell Bethe ansatz \cite{Bab94,Bab08}. Even
the most general integrable boundary condition (corresponding to the non-diagonal reflection matrix) was
obtained in 1999 \cite{Lim99}, however,  an interesting issue of
the exact solution for generic integrable boundary conditions is still
left for this model.

In this paper, we study the exactly solvable $t$-$J$ model with
generic integrable boundary fields. The Hamiltonian we shall
consider is
\begin{eqnarray}
    H&=&-t\sum_{\alpha,j=1}^{N-1}\mathcal{P}[C_{j,\alpha}^\dagger C_{j+1,\alpha}+C_{j+1,\alpha}^\dagger C_{j,\alpha}]\mathcal{P}\no\\
    &&+2t\sum_{j=1}^{N-1}[\vec{S}_j\cdot\vec{S}_{j+1}-\frac{1}{4}n_jn_{j+1}]+\xi_1n_1\no\\
    &&+2\vec{h}_1\cdot\vec{S}_1+\xi_Nn_N+2\vec{h}_N\cdot\vec{S}_N,\label{Ham}
\end{eqnarray}
where $N$ is the total number of sites; $t$ is the hopping constant;
$\mathcal{P}$ projects out double occupancies;
$C_{j,\alpha}^\dagger$ and $C_{j,\alpha}$ are the creation and
annihilation operators of electrons on the $j$-th site with spin component
$\alpha=\uparrow,\downarrow$;
$\vec{S}_j=\frac{1}{2}\sum_{\alpha,\beta}C_{j,\alpha}^\dagger\vec{\sigma}_{\alpha,\beta}C_{j,\beta}$
are the spin operators and $\vec{\sigma}$ are the Pauli matrices; $n_{j,\alpha}$ are
particle number operators; ${\vec h}_1=(h_1^x,h_1^y,h^z_1)$ and
${\vec h}_N=(h_N^x,h_N^y,h^z_N)$ are the boundary fields; $\xi_1$
and $\xi_N$ are the boundary chemical potentials. In the following
paper we shall show that for the proper choices of the boundary chemical
potentials, the model is exactly solvable for arbitrary boundary
magnetic fields and give the exact solution of the model.

The paper is organized as follows. In Section 2, we use
the coordinate Bethe ansatz method to derive the corresponding  the two-body scattering matrix (or R-matrix)
\cite{yang} and the reflection matrices (or  K-matrices) due to the boundary interaction. In Section 3, we
transform the eigenvalue problem into that of the inhomogeneous XXX spin
chain  with boundary fields, which allows us to apply the recently proposed
off-diagonal Bethe ansatz method \cite{cysw1,cysw2,cysw3} to solve it. The
exact spectrum of the Hamiltonian and the Bethe ansatz equations are
thus obtained. Section 4 is attributed to the reduction to the parallel or anti-parallel
boundary case. As an application of our solution, the surface energy of the model with
parallel fields are given in Section 5.  Concluding remarks are given in Section 6.

\section{Coordinate Bethe ansatz}
\label{CBA} \setcounter{equation}{0}
Due to the fact that the total number of electrons  of the super-symmetric t-J model is
conserved, we construct the eigenstate of the Hamiltonian  (\ref{Ham}) as
follows:
\begin{eqnarray}
|\Psi\rangle&=&\sum_{j=1}^M\sum_{\alpha_j=\uparrow,\downarrow}\sum_{x_j=1}^N\Psi^{\{\alpha\}}(x_1,\ldots,x_M)\no\\
&&\times C_{x_1,\alpha_1}^\dagger\cdots C_{x_M,\alpha_M}^\dagger|0\rangle,
\end{eqnarray}
where $\{\alpha\}=(\alpha_1,\ldots,\alpha_M)$ and $M$ is the total number
of electrons. To exclude double occupancy, we need to impose the following condition on the wave function
\begin{eqnarray}
\Psi^{\{\alpha\}}(\ldots,x,\ldots,x,\ldots)\equiv0.\label{psi-constraint}
\end{eqnarray}
This eigenvalue equation can be rewritten as
\begin{eqnarray}
&&-t\sum_{j=1}^M[(1-\delta_{x_j,N})\Psi^{\{\alpha\}}(\ldots,x_j+1,\ldots)+(1-\delta_{x_j,1})\Psi^{\{\alpha\}}(\ldots,x_j-1,\ldots)]\no\\
&&+\sum_{j=1}^M\sum_{\beta_j=\uparrow,\downarrow}}[\delta_{x_j,1}{(\xi_1+\vec
h}_1\cdot{\vec{\sigma}_{\alpha_j,\beta_j})+\delta_{x_j,N}(\xi_N+{\vec h}_N\cdot{\vec\sigma_{\alpha_j,\beta_j}})]\Psi^{\{\alpha\}_j}(x_1,\ldots,x_M)\nonumber\\
&&-t\sum_{j=1}^{N-1}\sum_{l=1}^{M}\sum_{k\neq{l}}^M\delta_{x_l,j}\delta_{x_k,j+1}\delta_{\alpha_l,-\alpha_k}[\Psi^{\{\alpha\}}(\ldots,x_l,\ldots,x_k,\ldots)+\Psi^{\{\alpha\}}(\ldots,x_k,\ldots,x_l,\ldots)]\no\\
&&\quad\quad=E\,\Psi^{\{\alpha\}}(x_1,\ldots,x_M),\label{eigen1}
\end{eqnarray}
where $\{\alpha\}_j$ means $\alpha_j$ is replaced by $\beta_j$ in
the set $\{\alpha\}$. Suppose the  wave function taking the
following Bethe ansatz form \cite{yang}:
\begin{eqnarray}
\Psi^{\{\alpha\}}(x_1,\ldots,x_M)&&\hspace{-0.2truecm}=\sum_{P,Q,r}A_P^{\{\alpha\},r}(Q)\exp[i\sum_{j=1}^Mr_{P_j}k_{P_j}x_{Q_j}]\no\\
&&\quad\times\theta(x_{Q_1}<x_{Q_2}<\cdots<x_{Q_M}),
\label{Wave-function}
\end{eqnarray}
where $Q=(Q_1,\ldots,Q_M)$ and $P=(P_1,\ldots,P_M)$ are the
permutations of $(1,\ldots,M)$; $r=(r_1,\ldots,r_M)$ with $r_j=\pm$;
$\theta(x_1<\ldots<x_M)$ is the generalized step function. For $x_j
\neq 1,N$, $x_i\neq1,N$, and $|x_i-x_j|>1$, the corresponding
eigenvalue is
\begin{eqnarray}
E=-2t\sum_{j=1}^M\cos k_j.\label{Eigenvalues}
\end{eqnarray}
When two electrons occupy two adjacent sites, namely,
$x_{Q_j}=x_{Q_{j+1}}-1=x$ and $x\neq1,N$, the Schr{\"o}dinger
equation (\ref{eigen1}) becomes
\begin{eqnarray}
&&[1+e^{ir_{P_j}k_{P_j}+ir_{P_{j+1}}k_{P_{j+1}}}-(1-P_{j,j+1})e^{ir_{P_{j+1}}k_{P_{j+1}}}] A_P^r(Q)\no\\
&&\quad +[1+e^{ir_{P_j}k_{P_j}+ir_{P_{j+1}}k_{P_{j+1}}}-(1-P_{j,j+1})e^{ir_{P_j}k_{P_j}}]A_{P'}^{r'}(Q)=0,\label{eigen2}
\end{eqnarray}
with $P'=(\ldots,P_{j+1},P_j,\ldots)$ ,$Q'=(\ldots,Q_{j+1},Q_j,\ldots)$ and
$r'=(\ldots,r_{j+1},r_j,\ldots)$. It is remarked that we have omitted the
superscript $\{\alpha\}$ and treated $A_P^r(Q)$ as a column vector
in the spin space. For convenience, let us introduce the permutation
operators ${\bar P}_{i,j}$ and $P_{i,j}$ in the coordinate and spin
sectors, respectively.
\begin{eqnarray}
{\bar P}_{j,j+1}A_P^r(Q)=A_P^r(Q').
\end{eqnarray}
Since the wave function of fermions is completely antisymmetric, we
have
\begin{eqnarray}
-P_{j,j+1}A_P^r(Q)=A_P^r(Q'),
\end{eqnarray}
and
\begin{eqnarray}
A_{P}^{r}(Q)=S(r_{P_j}\lambda_{P_j},r_{P_{j+1}}\lambda_{P_{j+1}})A_{P'}^{r'}(Q').\label{S}
\end{eqnarray}
After introducing a new parametrization
\begin{eqnarray}
e^{ik_j}=\frac{\lambda_j-\frac{i}{2}}{\lambda_j+\frac{i}{2}},\label{exchange}
\end{eqnarray}
we obtain the following $S-$matrix
\begin{eqnarray}
S(\lambda_j,\lambda_k)=S(\lambda_j-\lambda_k)=\frac{\lambda_j-\lambda_k+iP_{j,k}}{\lambda_j-\lambda_k+i}.
\end{eqnarray}
The $S-$matrix possesses the following property:
\begin{eqnarray}
S(\lambda)^{-1}=S(-\lambda).\no
\end{eqnarray}
Now we consider the case of $x_{Q1}=1$, $x_{Q_2}\neq2$. In this
case, the eigenvalue equation becomes
\begin{eqnarray}
&&-t\Psi^{\{\alpha\}}(2,\ldots)+\sum_{\beta_1}(\xi_1+{\vec
h}_1\cdot{\vec
\sigma}_{\alpha_1,\beta_1})\Psi^{\{\alpha\}_1}(1,\ldots)\no\\
&&\quad\quad=-2t\cos k_{P_1}\Psi^{\{\alpha\}}(1,\ldots).
\end{eqnarray}
This induces
\begin{eqnarray}
&&\sum_{\beta_1}(\xi_1+{\vec h}_1\cdot{\vec
\sigma}_{\alpha_1,\beta_1})\Psi^{\{\alpha\}_1}(1,\ldots)\no\\
&&\quad\quad=-t\Psi^{\{\alpha\}}(0,\ldots),\label{1,3}
\end{eqnarray}
or
\begin{eqnarray}
{A_P^{(+,\ldots)}(Q)}&&=-[t+(\xi_1+{\vec h}_1\cdot{\vec
\sigma_{P_1}})e^{ik_{P_1}}]^{-1}\no\\
&&\quad\times[t+(\xi_1+{\vec h}_1\cdot{\vec
\sigma_{P_1}})e^{-ik_{P_1}}]A^{(-,\ldots)}_P(Q)\no\\
&&={\bar K}_{P_1}^+(k_{P_1})A_P^{(-,\ldots)}(Q).\label{k+}
\end{eqnarray}
With the identity
\begin{eqnarray}
({\vec h}_1\cdot{\vec \sigma})^2={\vec h}_1^2,
\end{eqnarray}
one readily obtains
\begin{eqnarray}
\hspace{-0.1truecm}\bar{K}^+(k)=-\frac{t^2\hspace{-0.1truecm}+\hspace{-0.1truecm}\xi_1^2\hspace{-0.1truecm}-\hspace{-0.1truecm}\vec{h}_1^2\hspace{-0.1truecm}+\hspace{-0.1truecm}2\xi_1t\cos{k}\hspace{-0.1truecm}-\hspace{-0.1truecm}2it\sin{k}\,\vec{h}_1\cdot\vec{\sigma}}{(t+\xi_1e^{ik})^2-\vec{h}_1^2e^{2ik}},
\end{eqnarray}
and in terms of the new parameters $\lambda$ (\ref{exchange}) the $K$-matrix is given by
\begin{eqnarray}
\bar{K}^+(\lambda)&=&-\frac{\lambda+\frac{i}{2}}{\lambda-\frac{i}{2}}
\left[[(t+\xi_1)\lambda+\frac{i}{2}(t-\xi_1)]^2 -(\lambda-\frac{i}{2})^2\vec{h}_1^2\right]^{-1}\no\\
&&\times\left[(t^2+\xi_1^2-\vec{h}^2_1)(\lambda^2+\frac{1}{4})
 +2\xi_1t(\lambda^2-\frac{1}{4})+2i\lambda
t\,\vec{h}_1\cdot{\vec{\sigma}}\right].
\end{eqnarray}
For the case of $x_{Q_M}=N$, $x_{Q_{M-1}}\neq {N-1}$, the eigenvalue
equation is
\begin{eqnarray}
&&-t\Psi^{\{\alpha\}}(\ldots,{N\hspace{-0.1truecm}-\hspace{-0.1truecm}1})
\hspace{-0.1truecm}+\hspace{-0.1truecm}\sum_{\beta_M}(\xi_N\hspace{-0.1truecm}+\hspace{-0.1truecm}{\vec
h}_N\cdot{\vec
\sigma}_{\alpha_M,\beta_M})\Psi^{\{\alpha\}_M}\hspace{-0.08truecm}(\ldots,N)\no\\
&&\quad\quad =-2t\cos k_{P_M}\Psi^{\{\alpha\}}(\ldots,N).
\end{eqnarray}
This induces
\begin{eqnarray}
&&\sum_{\beta_M}(\xi_N+{\vec h}_N\cdot{\vec
\sigma}_{\alpha_M,\beta_M})\Psi^{\{\alpha\}_M}(\ldots,N)
=-t\Psi^{\{\alpha\}}(\ldots,N+1),
\end{eqnarray}
or
\begin{eqnarray}
A^{(\ldots,-)}_P(Q)&=&-e^{2ik_{P_M}N}[te^{-ik_{P_M}}+(\xi_N+{\vec h}_N\cdot{\vec\sigma_{P_M}})]^{-1}\no\\
&& \times[te^{ik_{P_M}}+(\xi_N+{\vec h}_N\cdot{\vec\sigma_{P_M}}))]A^{(\ldots,+)}_P(Q)\nonumber\\
&=&e^{2ik_{P_M}N}{\bar
K}_{P_M}^-(k_{P_M})A^{(\ldots,+)}_P(Q),\label{k-}
\end{eqnarray}
with
\begin{eqnarray}
\bar{K}^-\hspace{-0.05truecm}(k)\hspace{-0.1truecm}=\hspace{-0.1truecm}-\frac{t^2\hspace{-0.1truecm}+\hspace{-0.1truecm}\xi_N^2\hspace{-0.1truecm}-\hspace{-0.1truecm}\vec{h}_N^2\hspace{-0.1truecm}+\hspace{-0.1truecm}2\xi_Nt\cos{k}\hspace{-0.1truecm}-\hspace{-0.1truecm}2it\sin{k}\vec{h}_N\cdot\vec{\sigma}}{(te^{-ik}+\xi_N)^2-\vec{h}_N^2},\label{k-2}
\end{eqnarray}
and in terms of the new parameter $\lambda$ (\ref{exchange}) it reads
\begin{eqnarray}
\bar{K}^-(\lambda)&=&-\frac{\lambda-\frac{i}{2}}{\lambda+\frac{i}{2}}
\left[[(t+\xi_N)\lambda+\frac{i}{2}(t-\xi_N)]^2
-(\lambda-\frac{i}{2})^2\vec{h}_N^2\right]\no\\
&&\times \left[(t^2+\xi_N^2-\vec{h}^2_N)(\lambda^2+\frac{1}{4})
 +2\xi_Nt(\lambda^2-\frac{1}{4})+2i\lambda
t\,\vec{h}_N\cdot{\vec{\sigma}}\right].\label{k-3}
\end{eqnarray}
For the cases of $x_{Q_1}=1$, $x_{Q_2}=2$ and $x_{Q_{M-1}}=N-1$,
$x_{Q_M}=N$, one may check that (\ref{eigen1}) is self-consistent
with the solutions of $S$-matrix and $K$-matrices. The $S$-matrix
and $K$-matrices allows us to construct the following relations
\begin{eqnarray}
A^{(\ldots,+,\ldots)}&\hspace{-0.1truecm}=\hspace{-0.1truecm}&S_{j-1,j}(k_{j-1},k_j)\cdots S_{1,j}(k_1,k_j)A^{(+,\ldots)},\nonumber\\
A^{(+,\ldots)}&\hspace{-0.1truecm}=\hspace{-0.1truecm}&\bar{K}^+_j(k_j)\,A^{(-,\ldots)},\no\\
A^{(-,\ldots)}&\hspace{-0.1truecm}=\hspace{-0.1truecm}&S_{j,1}(-k_{j},k_1)\cdots S_{j,j-1}(-k_{j},k_{j-1})\no\\
&&\hspace{-0.1truecm}\times S_{j,j+1}(\hspace{-0.05truecm}-k_{j},k_{j+1})\cdots S_{j,M}(\hspace{-0.05truecm}-k_{j},k_M)A^{(\ldots,-)},\no\\
A^{(\ldots,-)}&\hspace{-0.1truecm}=\hspace{-0.1truecm}&e^{2ik_jN}\bar{K}^-_j(k_j)\,A^{(\ldots,+)},\no\\
A^{(\ldots,+)}&\hspace{-0.1truecm}=\hspace{-0.1truecm}&S_{M,j}(k_M,k_j)\cdots S_{j+1,j}(k_{j+1},k_j)A^{(\ldots,+,\ldots)}.\no
\end{eqnarray}
In terms of the parameter $\lambda$ (\ref{exchange}), the above relations give rise to  the following eigenvalue problem:
\begin{eqnarray}
&&\bar{\tau}(\lambda_j)A^{(\ldots,+,\ldots)}=(\frac{2\lambda_j-i}{2\lambda_j+i})^{-2N}A^{(\ldots,+,\ldots)},\label{bar-tau}
\end{eqnarray}
with the resulting operator
\begin{eqnarray}
{\bar\tau}(u)&=&S_{j-1,j}(\lambda_{j-1},u)\cdots S_{1,j}(\lambda_{1},u){\bar K}_j^{+}(u)\no\\
&&\times S_{j,1}(-u,\lambda_1)\cdots S_{j,j-1}(-u,\lambda_{j-1})\no\\
&&\times S_{j,j+1}(-u,\lambda_{j+1})\cdots S_{j,M}(-u,\lambda_M) {\bar K}_j^-(u)\no\\
&&\times S_{M,j}(\lambda_M,u)
\cdots S_{j+1,j}(\lambda_{j+1},u)\no\\
&=&S_{j-1,j}(\lambda_{j-1}-u)\cdots S_{1,j}(\lambda_{1}-u){\bar K}_j^{+}(u)\no\\
&&\times S_{j,1}(-u-\lambda_1)\cdots
S_{j,j-1}(-u-\lambda_{j-1})\no\\
&&\times S_{j,j+1}(-u-\lambda_{j+1})\cdots S_{j,M}(-u-\lambda_M) {\bar K}_j^-(u)\no\\
&&\times S_{M,j}(\lambda_M-u)
\cdots S_{j+1,j}(\lambda_{j+1}-u).\label{H-operators}
\end{eqnarray}
To ensure the integrability of the model, namely, to assure that the resulting operator with different values of $u$
commute with each other: $[{\bar\tau}(u),{\bar\tau}(v)]=0$, the corresponding $K$-matrices $\bar{K}^{\pm}(u)$ have to satisfy   the following
reflection equation \cite{Skl88}
\begin{eqnarray}
&&S_{1,2}(u_1-u_2){\bar K}_1^\pm(u_1)S_{1,2}(u_1+u_2){\bar
K}_2^\pm(u_2)\no\\
&&\quad\quad={\bar K}_2^\pm(u_2)S_{1,2}(u_1+u_2){\bar
K}_1^\pm(u_1)S_{1,2}(u_1-u_2),
\end{eqnarray}
which induces the following integrable conditions of the model
\begin{eqnarray}
(t+\xi_1)^2=\vec{h}_1^2,\quad (t+\xi_N)^2=\vec{h}_N^2.
\end{eqnarray}
Under this restriction, the reflection matrices become
\begin{eqnarray}
&&\bar{K}^-(\lambda)=\frac{2\lambda-i}{2\lambda+i}\frac{\xi_N-2i\lambda\,\vec{h}_N\cdot{\vec{\sigma}}}{\xi_N+2i\lambda(t+\xi_N)},\\
&&\bar{K}^+(\lambda)=\frac{2\lambda+i}{2\lambda-i}\frac{\xi_1-2i\lambda\,\vec{h}_1\cdot{\vec{\sigma}}}{\xi_1+2i\lambda(t+\xi_1)}.
\end{eqnarray}
Similarly as that of the Hubbard model with
arbitrary boundary magnetic fields \cite{Li13}, in our case the eigenvalue problem (\ref{bar-tau}) can be
transformed into that of the transfer matrix of the inhomogeneous
XXX spin chain model with arbitrary boundary fields and thus can be solved via the off-diagonal
Bethe ansatz \cite{cysw1,cysw2,cysw3}.
\section{Off-diagonal Bethe ansatz}
\label{OBA} \setcounter{equation}{0}
Before going further, let us
introduce the following R-matrix and K-matrices:
\begin{eqnarray}
&&R_{0,j}(u)=u+\eta{P_{0,j}},\label{P}\\
&&K^-_0(u)=p+u{\vec{h}_N}\cdot{\vec{\sigma_0}},\label{K-}\\
&&K^+_0(u)=q-(u+\eta){\vec{h}_1}\cdot{\vec{\sigma_0}},\label{K+}
\end{eqnarray}
where
\begin{eqnarray}
\eta=i,\quad p=\frac{\xi_N}{2i}, \quad
q=-\frac{\xi_1}{2i}.\nonumber
\end{eqnarray}
We remark that the $K$-matrices (\ref{K-})-(\ref{K+}) are the most general reflection matrices associated the XXX spin chain \cite{Veg93,Gho94}.
The $R-$martrix has the following properties:
\begin{eqnarray}
&&\hspace{-0.6truecm}\mbox{Initial condition}:\,R_{1,2}(0)= \eta P_{1,2},\label{Intital}\\
&&\hspace{-0.6truecm}\mbox{Unitarity relation}:\,R_{1,2}(u)R_{1,2}(-u)=-(u\hspace{-0.05truecm}+\hspace{-0.05truecm}\eta)(u\hspace{-0.05truecm}-\hspace{-0.05truecm}\eta)\,{\rm id},\label{Unitarity}\\
&&\hspace{-0.6truecm}\mbox{Crossing relation}:\,R_{12}(u)\hspace{-0.1truecm}=\hspace{-0.1truecm}V_1R_{12}^{t_2}(-u\hspace{-0.1truecm}-\hspace{-0.1truecm}\eta)V_1,\,V=-i\sigma^y.\label{Crosing}
\end{eqnarray}
The following Yang-Baxter equation, the reflection equation and the dual reflection equation also hold:
\begin{eqnarray}
&&R_{0,0'}(u-v)R_{0,1}(u)R_{0',1}(v)=R_{0',1}(v)R_{0,1}(u)R_{0,0'}(u-v),\label{YBeq}\\
&&R_{0,0'}(u-v)K_0^-(u)R_{0,0'}(u+v)K_{0'}^-(v)=K_{0'}^-(v)R_{0,0'}(u+v)K_0^-(u)R_{0,0'}(u-v),\label{RE}\\
&&R_{0,0'}(v-u)K_0^+(u)R_{0,0'}(-u-v-2\eta)K_{0'}^+(v)\no\\
&&\quad\quad\quad\quad=K_{0'}^+(v)R_{0,0'}(-u-v-2\eta)K_0^+(u)R_{0,0'}(v-u).\label{DRE}
\end{eqnarray}
We introduce the inhomogeneous double-row monodromy matrix
\begin{eqnarray}
T_0(u)&=&R_{0,1}(u-\lambda_1)\cdots R_{0,M}(u-\lambda_M)K_0^-(u)\no\\
&&\times R_{M,0}(u+\lambda_M)\cdots{R_{1,0}}(u+\lambda_1),
\end{eqnarray}
and the associated transfer matrix $\tau(u)$ is given by \cite{Skl88}
\begin{eqnarray}
\tau(u)=tr_0\left\{{K}_0^+(u)T_0(u)\right\}.\label{tau}
\end{eqnarray}
From (\ref{YBeq}), (\ref{RE}) and (\ref{DRE}) one may derive
\begin{eqnarray}
&&R_{0,0'}(u-v)T_0(u)R_{0,0'}(u+v)T_{0'}(v)=T_{0'}(v)R_{0,0'}(u+v)T_0(u)R_{0,0'}(u-v),
\end{eqnarray}
and thus the transfer matrices with different spectrum parameters commute with each other,
\begin{eqnarray}
[\tau(u),\, \tau(v)]=0,
\end{eqnarray}
which ensures the integrability of the associated spin chain.
Let $u=-\lambda_j$, using  the initial condition (\ref{Intital}) and Yang-Baxter equation (\ref{YBeq})  we can express
the transfer matrix at special point in terms of the K-matrices and the R-matrix
\begin{eqnarray}
\tau(-\lambda_j)&=&R_{j-1,j}(\lambda_{j-1}-\lambda_j)\ldots R_{1,j}(\lambda_{1}-\lambda_j)\no\\
&&\times\,tr_0\left\{K^+_0(-\lambda_j)R_{0,j}(-2\lambda_j)R_{0,j}(0)\right\}\no\\
&&\times R_{j,1}(-\lambda_{j}-\lambda_1)\ldots R_{j,j-1}(-\lambda_{j}-\lambda_{j-1})\nonumber\\
&&\times R_{j,j+1}(-\lambda_{j}-\lambda_{j+1})
\ldots R_{j,M}(-\lambda_{j}-\lambda_M)\no\\
&&\times K^-_j(\lambda_j)\,R_{M,j}(\lambda_M\hspace{-0.1truecm}-\hspace{-0.1truecm}\lambda_j)\ldots R_{j+2,j}(\lambda_{j+2}\hspace{-0.1truecm}-\hspace{-0.1truecm}\lambda_j)\no\\
&&\times R_{j+1,j}(\lambda_{j+1}-\lambda_j).\label{tau-j}
\end{eqnarray}
Noticing that
\begin{eqnarray}
&&S_{j,l}(\lambda_j,\lambda_l)=\frac{R_{j,l}(\lambda_j-\lambda_l)}{\lambda_j-\lambda_l+\eta},\\
&&S_{j,l}(-\lambda_j,\lambda_l)=\frac{R_{j,l}(-\lambda_j-\lambda_l)}{-\lambda_j-\lambda_l+\eta},\\
&&\bar{K}^-_j(\lambda_j)=\frac{2\lambda_j-\eta}{2\lambda_j+\eta}\frac{K^-_j(-\lambda_j)}{p+\lambda_j(t+\xi_N)},\\
&&\bar{K}^+_j(\lambda_j)=-\frac{tr_0\left\{K^+_0(-\lambda_j)R_{0,j}(-2\lambda_j)
R_{0,j}(0)\right\}}{2\eta(\lambda_j-\eta)[q-\lambda_j(t+\xi_1)]}
\frac{2\lambda_j+\eta}{2\lambda_j-\eta},
\end{eqnarray}
we have the following important identification between the operators
$\{\bar{\tau}(\lambda_j)\}$ given by (\ref{H-operators}) appeared in the eigenvalue problem of the
super-symmetric t-J model with boundary fields  and  the transfer matrices $\{\tau(\lambda_j)\}$ of the open XXX spin chain with boundary fields
\begin{eqnarray}
\bar{\tau}(\lambda_j)&\hspace{-0.1truecm}=\hspace{-0.1truecm}&\prod_{l\neq{j}}^M(\lambda_j-\lambda_l-\eta)^{-1}(\lambda_j+\lambda_l-\eta)^{-1}\frac{1}{2\eta(\lambda_j-\eta)}\no\\
&&\times\frac{\tau(-\lambda_j)}{[p+\lambda_j(t+\xi_N)][-q+\lambda_j(t+\xi_1)]}.\label{relation}
\end{eqnarray}
The eigenvalue problem (\ref{bar-tau}) is thus equivalent to that of
diagonalizing the transfer matrix of the inhomogeneous open XXX chain model with boundary fields.
Here we naturally have the ``inhomogeneous" parameters
$\lambda_j$ which is related to the quasi-momentum (\ref{exchange}) of the electrons and the crossing parameter $\eta=i$. Thanks to the works \cite{cysw1,cysw2,cysw3},
the transfer matrix (\ref{tau}) of the open XXX chain with arbitrary boundary fields which is specified  by the K-matrices
$K^{\pm}(u)$ (\ref{K-}) and (\ref{K+}) can be exactly diagonalized by off-diagonal Bethe ansatz method. In the following, we
shall use the method in \cite{cysw3} to solve the eigenvalue problem (\ref{bar-tau}) of the super-symmetric t-J model with general boundary fields.
Suppose $|\Psi\rangle$ is an eigenstate of $\tau(u)$ and the corresponding eigenvalue is $\Lambda(u)$
\begin{eqnarray}
\tau(u)|\Psi\rangle=\Lambda(u)|\Psi\rangle.
\end{eqnarray}
Following the method in \cite{cysw3}, we find that $\Lambda(u)$ possesses the following  properties:
\begin{eqnarray}
&&\Lambda(u)=\Lambda(-u-\eta),\label{EIGENVALUE1} \\
&&\Lambda(0)=2pq\prod_{l=1}^M-(\lambda_l-\eta)(\lambda_l+\eta),\label{EIGENVALUE2}\\
&&\Lambda(u){\sim{-2}}\vec{h}_1\cdot{\vec{h}_N}u^{2M+2}+\cdots,u\rightarrow\pm\infty,\label{EIGENVALUE3}\\
&&\Lambda(\lambda_j)\Lambda(\lambda_j-\eta)\hspace{-0.1truecm}=\hspace{-0.1truecm}\frac{4(\lambda_j^2-\eta^2)}{4\lambda_j^2-\eta^2}(q^2-\lambda_j^2\vec{h}_1^2)(p^2-\lambda_j^2\vec{h}^2_N)\no\\
&&\qquad\qquad\times\prod_{l=1}^M[(\lambda_j\hspace{-0.1truecm}+\hspace{-0.1truecm}\lambda_l)^2\hspace{-0.1truecm}-\hspace{-0.1truecm}\eta^2][(\lambda_j\hspace{-0.1truecm}-\hspace{-0.1truecm}\lambda_l)^2
\hspace{-0.1truecm}-\hspace{-0.1truecm}\eta^2],\,\, j=1,2,3\ldots,M.\label{EIGENVALUE4}
\end{eqnarray}
Moreover, the explicit expressions (\ref{P})-(\ref{K+}) of the R-matrix and K-matrices implies that $\Lambda(u)$, as a function of $u$,
is a polynomial of degree $2M+2$, hence $\Lambda(u)$
can be completely determined by
(\ref{EIGENVALUE1})-(\ref{EIGENVALUE4}).

For convenience, we introduce the following notations
\begin{eqnarray}
&&A(u)=\prod^M_{l=1}(u-\lambda_l+\eta)(u+\lambda_l+\eta),\label{A}\\
&&a(u)=\frac{2u+2\eta}{2u+\eta}[p+u{\rm sgn}(\vec{h}_1\cdot{\vec{h}_N})|\vec{h}_N|]\no\\
&&\quad\quad\quad\quad\times(q-u|\vec{h}_1|)A(u),\label{a}\\
&&d(u)=a(-u-\eta),\label{d}\\
&&c=2[{\rm sgn}(\vec{h}_1\cdot{\vec{h}_N})|\vec{h}_1||\vec{h}_N|-\vec{h}_1\cdot{\vec{h}_N}].\label{c}
\end{eqnarray}
\subsection{Even $M$ case}

As in \cite{cysw3}, we make the following functional $T-Q$ ansatz
for an even $M$:
\begin{eqnarray}
&&\Lambda(u)=a(u)\frac{Q_1(u-\eta)}{Q_2(u)}+d(u)\frac{Q_2(u+\eta)}{Q_1(u)}
+cu(u+\eta)\frac{A(u)A(-u-\eta)}{Q_1(u)Q_2(u)},\label{Lambda1}
\end{eqnarray}
where the functions $Q_1(u)$ and $ Q_2(u)$ are parameterized by $M$ Bethe roots $\{\mu_j|j=1,\ldots,M)\}$ for a generic non-vanishing $c$ as follows
\begin{eqnarray}
&&Q_1(u)=\prod_{j=1}^M(u-\mu_j),\quad Q_2(u)=\prod_{j=1}^M(u+\mu_j+\eta)=Q_1(-u-\eta).
\end{eqnarray}
$\Lambda(u)$ becomes the eigenvalue of the transfer matrix $\tau(u)$ if the parameters $\{\mu_j|j=1,\ldots,M\}$ satisfies the following Bethe ansatz equations:
\begin{eqnarray}
&&\hspace{-1.2truecm}\frac{c\,(\mu_j+\eta)(2\mu_j+\eta)}{2[p-(\mu_j+\eta)\,{\rm sgn}(\vec h_1\cdot\vec h_N)|\vec h_N|][q+(\mu_j+\eta)|\vec h_1|]}\no\\
&&\hspace{-1.2truecm}=-\prod_{l=1}^M\frac{(\mu_j+\mu_l+\eta)(\mu_j+\mu_l+2\eta)}{(\mu_j-\lambda_l+\eta)(\mu_j+\lambda_l+\eta)}, j=1,\ldots,M,\label{BA-even-1}
\end{eqnarray}
With the identification (\ref{relation}), we get the other Bethe
ansatz equations
\begin{eqnarray}
&&\frac{[p-\lambda_j{\rm sgn}(\vec{h}_1\cdot{\vec{h}_N})|\vec{h}_N|](q+\lambda_j|\vec{h}_1|)}{[p+\lambda_j(t+\xi_N)][q-\lambda_j(t+\xi_1)]}(\frac{2\lambda_j-\eta}{2\lambda_j+\eta})^{2N}\no\\
&&\quad\quad=\prod^M_{l=1}\frac{\lambda_j-\mu_l-\eta}{\lambda_j+\mu_l+\eta},\quad j=1,\ldots,M.\label{BA-even-2}
\end{eqnarray}
Then from the solutions of the Bethe ansatz equations (\ref{BA-even-1}) and (\ref{BA-even-2}), one can reconstruct the exact wave
functions (\ref{Wave-function}) with even number of electrons for the super-symmetric t-J model with boundary fields, the corresponding eigenvalues
are given by (\ref{Eigenvalues}).

\subsection{Odd $M$ case}
For an odd $N$, we make the following functional $T-Q$ ansatz
\begin{eqnarray}
&&\Lambda(u)= a(u)\frac{Q_1(u-\eta)}{Q_2(u)}+ d(u)\frac{Q_2(u+\eta)}{Q_1(u)}
+ cu^2(u+\eta)^2\frac{A(u) A(-u-\eta)}{Q_1(u)Q_2(u)},\label{Lambda2}
\end{eqnarray}
where the functions $a(u)$, $d(u)$ and $A(u)$ and the parameter $c$ are given by (\ref{A})-(\ref{c}) respectively. The functions $Q_1(u)$ and $Q_2(u)$ are parameterized by $M+1$ Bethe roots $\{\mu_j|j=1,\ldots,M+1\}$ for a generic non-vanishing $c$ as follows:
\begin{eqnarray}
&&Q_1(u)=\prod_{j=1}^{M+1}(u-\mu_j),\quad Q_2(u)=\prod_{j=1}^{M+1}(u+\mu_j+\eta)=Q_1(-u-\eta).
\end{eqnarray}
The $M$ quasi-momentum $\{k_j\}$ (or $\{\lambda_j\}$)  and the $M+1$ parameters
$\{\mu_j|j=1,\ldots,M+1\}$ need to satisfy the following Bethe ansatz equations:
\begin{eqnarray}
\frac{[p-\lambda_j {\rm sgn}(\vec{h}_1\cdot{\vec{h}_N})|\vec{h}_N|](q+\lambda_j|\vec{h}_1|)}{[p+\lambda_j(t+\xi_N)][q-\lambda_j(t+\xi_1)]}
(\frac{2\lambda_j-\eta}{2\lambda_j+\eta})^{2N}
=\prod_{l=1}^{M+1}\frac{\lambda_j-\mu_l-\eta}{\lambda_j+\mu_l+\eta},\quad j=1,\ldots,M,\label{BA-odd-1}
\end{eqnarray}
\begin{eqnarray}
&&\frac{c\mu_j(\mu_j+\eta)^2(2\mu_j+\eta)}{2[p-(\mu_j+\eta){\rm sgn}(\vec{h}_1\cdot{\vec{h}}_N)|\vec{h}_N|][q+(\mu_j+\eta)|\vec{h}_1|]}\nonumber\\
&&\quad\quad=-\frac{\prod_{l=1}^{M+1}(\mu_j+\mu_l+\eta)(\mu_j+\mu_l+2\eta)}{\prod_{l=1}^M(\mu_j-\lambda_l+\eta)(\mu_j+\lambda_l+\eta)},\quad
j=1,\ldots,M+1.\label{BA-odd-2}
\end{eqnarray}
Then from the solutions of the Bethe ansatz equations (\ref{BA-odd-1}) and (\ref{BA-odd-2}), one can reconstruct the exact wave
functions (\ref{Wave-function}) with odd number of electrons for the super-symmetric t-J model with boundary fields, the corresponding eigenvalues
are given by (\ref{Eigenvalues})
\section{Reduction to the parallel boundary case}
\label{RED} \setcounter{equation}{0}

When the two boundary fields $\vec{h}_1$ and $\vec{h}_N$ are parallel or anti-parallel, the parameter $c$ is vanishing. The resulting $T-Q$ relation becomes the conventional one
\begin{eqnarray}
\Lambda(u)=a(u)\frac{Q(u-\eta)}{Q(u)}+d(u)\frac{Q(u+\eta)}{Q(u)},\label{Lambda3}
\end{eqnarray}
with
\begin{eqnarray}
Q(u)&=&\prod_{l=1}^m(u-\gamma_l)(u+\gamma_l+\eta)=Q(-u-\eta),\quad
m=0,1,\cdots,M.
\end{eqnarray}
Parameter $\{\gamma_j\}$ and quasi-momentum $\{\lambda_j\}$ satisfy the following Bethe ansatz equations:
\begin{eqnarray}
&&\frac{[p-\lambda_j{\rm sgn}(\vec h_1\cdot\vec h_N)|\vec{h}_N|](q+\lambda_j|\vec{h}_1|)}{[p+\lambda_j(t+\xi_N)][q-\lambda_j(t+\xi_1)]}
(\frac{2\lambda_j-\eta}{2\lambda_j+\eta})^{2N}\no\\
&&\quad\quad=\prod_{l=1}^m\frac{(\lambda_j-\gamma_l-\eta)(\lambda_j+\gamma_l)}{(\lambda_j+\gamma_l+\eta)(\lambda_j-\gamma_l)},\quad j=1,\ldots,M,\label{BA-3-2}
\end{eqnarray}
\begin{eqnarray}
&&\frac{\gamma_j[p-(\gamma_j+\eta){\rm sgn}(\vec h_1\cdot\vec h_N)|\vec{h}_N|][q+(\gamma_j+\eta)|\vec{h}_1|]}{(\gamma_j+\eta)[p+\gamma_j{\rm sgn}(\vec h_1\cdot\vec h_N)|\vec{h}_N|](q-\gamma_j|\vec{h}_1|)}\no\\
&&\quad\quad\quad\quad\times\prod_{l=1}^M\frac{(\gamma_j+\lambda_l)(\gamma_j-\lambda_l)}{(\gamma_j-\lambda_l+\eta)(\gamma_j+\lambda_l+\eta)}\nonumber\\
&&\quad\quad=-\prod_{l=1}^m\frac{(\gamma_j\hspace{-0.1truecm}-\hspace{-0.1truecm}\gamma_l\hspace{-0.1truecm}-\hspace{-0.1truecm}\eta)
(\gamma_j\hspace{-0.1truecm}+\hspace{-0.1truecm}\gamma_l)}{(\gamma_j\hspace{-0.1truecm}-\hspace{-0.1truecm}\gamma_l\hspace{-0.1truecm}+\hspace{-0.1truecm}\eta)(\gamma_j\hspace{-0.1truecm}+\hspace{-0.1truecm}\gamma_l\hspace{-0.1truecm}+\hspace{-0.1truecm}2\eta)},\quad j=1,\ldots,m.\label{BA-3-3}
\end{eqnarray}
Then from the solutions of the Bethe ansatz equations (\ref{BA-3-2}) and (\ref{BA-3-3}), one can reconstruct the exact wave
functions (\ref{Wave-function})  for the super-symmetric t-J model with parallel or anti-parallel boundary fields, the corresponding eigenvalues
are given by (\ref{Eigenvalues}).

\section{Surface energy}
\label{SE} \setcounter{equation}{0}

As an application of our exact solution of the supersymmetry t-J
model with boundary fields, here we study the surface energy of the
supersymmetry t-J model with parallel boundary fields \footnote{The
generalization to the case of  the generic non-diagonal boundary is
nontrivial even for the XXX open spin chain case \cite{Jia13,Nep13}.}. In the interesting paper
\cite{Ess96}, Essler calculated the surface energy for this particular case. Here we only list the
main results and we refer the reader to \cite{Ess96} for more details. Let
us introduce a new parameter
\begin{eqnarray}
\theta_j=\gamma_j+\frac{\eta}{2}.
\end{eqnarray}

\subsection{The case of $t+\xi_N=-|\vec h_N|$ and $t+\xi_1=-|\vec h_1|$}

In this case, the functions (\ref{BA-3-2}) and (\ref{BA-3-3}) become
\begin{eqnarray}
&&(\frac{\lambda_j-\frac{\eta}{2}}{\lambda_j+\frac{\eta}{2}})^{2N}=\prod_{l=1}^m\frac{\lambda_j-\theta_l-\frac{\eta}{2}}{\lambda_j-\theta_l+\frac{\eta}{2}}
\frac{\lambda_j+\theta_l-\frac{\eta}{2}}{\lambda_j+\theta_l+\frac{\eta}{2}},\quad
 j=1,2,\ldots,M,\label{BA-4-1}
\end{eqnarray}
and
\begin{eqnarray}
&&\frac{\theta_j-\frac{\eta}{2}}{\theta_j+\frac{\eta}{2}}\frac{\theta_j-c\eta}{\theta_j+c\eta}\frac{\theta_j-d\eta}{\theta_j+d\eta}
\prod_{l=1}^M\frac{\theta_j+\lambda_l-\frac{\eta}{2}}{\theta_j+\lambda_l+\frac{\eta}{2}}\frac{\theta_j-\lambda_l-\frac{\eta}{2}}
{\theta_j-\lambda_l+\frac{\eta}{2}}\no\\
&&\quad\quad=-\prod_{l=1}^m\frac{\theta_j-\theta_l-\eta}{\theta_j-\theta_l+\eta}\frac{\theta_j+\theta_l-\eta}{\theta_j+\theta_l+\eta},\quad  j=1,2,\ldots,m,\label{BA-4-2}
\end{eqnarray}
where
\begin{eqnarray}
c=-\frac{\xi_N}{2|\vec{h}_N|}-\frac{1}{2},\quad d=-\frac{\xi_1}{2|\vec{h}_1|}-\frac{1}{2}.\label{c,d}
\end{eqnarray}
The logarithm of equations (\ref{BA-4-1}) and (\ref{BA-4-2}) are
\begin{eqnarray}
2N\ln{\frac{\lambda_j-\frac{\eta}{2}}{\lambda_j+\frac{\eta}{2}}}=
\sum_{l=1}^m\lt\{\ln{\frac{\lambda_j-\theta_l-\frac{\eta}{2}}{\lambda_j-\theta_l+\frac{\eta}{2}}}
+\ln{\frac{\lambda_j+\theta_l-\frac{\eta}{2}}{\lambda_j+\theta_l+\frac{\eta}{2}}}\rt\}
+2\pi\eta I_j,\, j=1,2,\ldots,M,\label{BA-4-3}
\end{eqnarray}
and
\begin{eqnarray}
&&\ln{\frac{\theta_j-\frac{\eta}{2}}{\theta_j+\frac{\eta}{2}}}
+\ln{\frac{\theta_j-c\eta}{\theta_j+c\eta}}+\ln{\frac{\theta_j-d\eta}{\theta_j+d\eta}}
+\sum_{l=1}^M\lt\{\ln{\frac{\theta_j+\lambda_l-\frac{\eta}{2}}{\theta_j+\lambda_l+\frac{\eta}{2}}}
+\ln{\frac{\theta_j-\lambda_l-\frac{\eta}{2}}{\theta_j-\lambda_l+\frac{\eta}{2}}}\rt\}\no\\
&&\quad\quad =\pi\eta+\sum_{l=1}^m\lt\{\ln{\frac{\theta_j-\theta_l-\eta}{\theta_j-\theta_l+\eta}}
+\ln{\frac{\theta_j+\theta_l-\eta}{\theta_j+\theta_l+\eta}}\rt\}+2\pi\eta I_j',
\,j=1,2,\ldots,m,\label{BA-4-4}
\end{eqnarray}
where $I_j$ and $I_j'$ are integers. In the thermodynamic limit $\{\lambda_j\}$ and $\{\theta_j\}$ distribute with densities $\rho(\lambda)$ and
$\sigma(\theta)$ respectively. Due to the fact that $\lambda_j=0$ and $\theta_j=0$ are the solution of (\ref{BA-4-1}) and (\ref{BA-4-2}) and they make
the wave-function vanishing, they should be excluded, namely, the densities corresponding to the ground state in thermodynamic limit are \cite{Alc87,Kap96}
\begin{eqnarray}
\rho(\lambda)=\frac{1}{2N}\frac{dI}{d\lambda}\hspace{-0.02truecm}-\hspace{-0.02truecm}
\frac{1}{2N}\delta(\lambda),\quad
\sigma(\theta)\hspace{-0.02truecm}=\hspace{-0.02truecm}\frac{1}{2N}\frac{dI'}{d\theta}\hspace{-0.02truecm}-\hspace{-0.02truecm}\frac{1}{2N}\delta(\theta).
\end{eqnarray}
and $\rho(\lambda)=\rho(-\lambda)$, $\sigma(\theta)=\sigma(-\theta)$.

Let us firstly consider  the case: $t>0$, which is corresponding to $c>0$ and $d>0$. Taking the derivative of (\ref{BA-4-3}) and (\ref{BA-4-4}), we have
\begin{eqnarray}
a_1(\lambda)-\frac{1}{2N}\delta(\lambda)=\int_{-\infty}^{\infty}a_1(\lambda-x)\sigma(x)dx+\rho(\lambda),\label{BA-4-5}
\end{eqnarray}
and
\begin{eqnarray}
&&\frac{1}{2N}[a_1(\theta)+a_{2c}(\theta)+a_{2d}({\theta})]+\int_{-\infty}^{\infty}a_1(\theta-y)\rho(y)dy\no\\
&&\quad\quad=\int_{-\infty}^{\infty}a_2(\theta-y)\sigma(\theta)dy+\sigma(\theta)+\frac{1}{2N}\delta(\theta).\label{BA-4-6}
\end{eqnarray}
Here the function $a_n(z)$ is defined by
\begin{eqnarray}
a_n(z)=\frac{1}{2\pi}\frac{n}{z^2+\frac{n^2}{4}},n>0.
\end{eqnarray}
Using the Fourier expansion
\begin{eqnarray}
&&\tilde{f}(\omega)=\int_{-\infty}^{\infty} e^{i\omega x}f(x)dx,\quad
f(x)=\frac{1}{2\pi}\int_{-\infty}^{\infty} e^{-i\omega x}\tilde{f}(\omega)d\omega,\label{F.E.}
\end{eqnarray}
(\ref{BA-4-5}) and (\ref{BA-4-6}) become
\begin{eqnarray}
\tilde{a}_1(\omega)-\frac{1}{2N}=\tilde{a}_1(\omega)\tilde{\sigma}(\omega)+\tilde{\rho}(\omega),\label{BA-4-7}
\end{eqnarray}
and
\begin{eqnarray}
&&\frac{1}{2N}[\tilde{a}_1(\omega)+\tilde{a}_{2c}(\omega)+\tilde{a}_{2d}(\omega)-1]
=[\tilde{a}_2(\omega)+1]\tilde{\sigma}(\omega)-\tilde{a}_1(\omega)\tilde{\rho}(\omega),\label{BA-4-8}
\end{eqnarray}
where
\begin{eqnarray}
\tilde{a}_n(\omega)=e^{-\frac{n}{2}|\omega|},\quad
n>0.\label{a-1210}
\end{eqnarray}
Solving equations (\ref{BA-4-7}) and (\ref{BA-4-8}) yields that
\begin{eqnarray}
\tilde{\rho}(\omega)&=&\frac{1}{2\tilde{a}_2(\omega)+1}[(\tilde{a}_2(\omega)+1)\tilde{a}_1(\omega)-C(\omega)],\label{density}\\
C(\omega)&=&\frac{1}{2N}[\tilde{a}_2(\omega)+1+\tilde{a}_1(\omega)(\tilde{a}_1(\omega)+\tilde{a}_{2c}(\omega)
+\tilde{a}_{2d}(\omega)-1)].\label{c-function}
\end{eqnarray}
For the periodic t-J model, $C(\omega)$ is vanishing, the number of electrons in the ground state is
\begin{eqnarray}
M&\hspace{-0.1truecm}=\hspace{-0.1truecm}&N\int_{-\infty}^{\infty}\rho(\lambda)d\lambda=\frac{N}{2\pi}\int_{-\infty}^{\infty}\int_{-\infty}^{\infty}\tilde{\rho}(\omega)e^{-i\omega\lambda} d\omega d\lambda\no\\
&=&N\tilde{\rho}(0)=\frac{2}{3}N
\end{eqnarray}
and the ground state energy in the thermodynamic limit is
\begin{eqnarray}
E_g^0&=&-2t\sum_{j=1}^M\frac{\lambda_j^2-\frac{1}{4}}{\lambda_j^2+\frac{1}{4}}=-2Nt\int_{-\infty}^{\infty}\rho(\lambda)\frac{\lambda^2-\frac{1}{4}}{\lambda^2+\frac{1}{4}}d\lambda\no\\
&=&-2Nt\int_{-\infty}^{\infty}\tilde{\rho}(\omega)[\delta(\omega)-\frac{1}{2}\tilde{a}_1(\omega)]d\omega\no\\
&=&-\frac{1}{3}Nt+\frac{\ln3}{2}Nt.\label{E0-1}
\end{eqnarray}
For the t-J model with open boundary condition, we find that there is no boundary strings and that  $C(w)$ given by (\ref{c-function}) does not vanish.  Then the ground state energy in the thermodynamic limit is
\begin{eqnarray}
E_g&=&-2t\sum_{j=1}^M\frac{\lambda_j^2-\frac{1}{4}}{\lambda_j^2+\frac{1}{4}}=-2Nt\int_{-\infty}^{\infty}\rho(\lambda)\frac{\lambda^2-\frac{1}{4}}{\lambda^2+\frac{1}{4}}d\lambda\no\\
&=&-2Nt\int_{-\infty}^{\infty}\tilde{\rho}(\omega)[\delta(\omega)-\frac{1}{2}\tilde{a}_1(\omega)]d\omega\no\\
&=&E_g^0+2Nt\int_{-\infty}^{\infty}\frac{C(\omega)}{2\tilde{a}_2(\omega)+1}[\delta(\omega)-\frac{1}{2}\tilde{a}_1(\omega)]d\omega\no\\
&=&E_g^0+\frac{\ln3}{2}t-\frac{2}{3}t-B_ct-B_dt,\label{E0-2}
\end{eqnarray}
with
\begin{eqnarray}
B_{p>-1}&=&\int_0^1\frac{x^p}{2x+1}dx
=\int_0^{\frac{1}{2}}\frac{x^p}{2x+1}dx+\int_{\frac{1}{2}}^1\frac{1}{2}\frac{x^{p-1}}{1+\frac{1}{2x}}dx\no\\
&=&\int_0^{\frac{1}{2}}\sum_{n=0}^{\infty}(-1)^n2^nx^{n+p}dx
+\int_{\frac{1}{2}}^1\sum_{n=0}^{\infty}(-1)^n\frac{1}{2^{n+1}}x^{p-1-n}dx\no\\
&=&\sum_{n=0}^{\infty}(-1)^n\frac{1}{2^{p+1}}\frac{1}{p+n+1}+\sum_{n=0}^{\infty}(1-\delta_{p,n})
(-1)^n(\frac{1}{2^{n+1}}-\frac{1}{2^{p+1}})\frac{1}{p-n}\no\\
&&+\delta_{p,n}(-1)^p\frac{1}{2^{p+1}}\ln2.\label{B}
\end{eqnarray}
For the case of $t<0$, which is corresponding to $c<0$ and $d<0$ (\ref{E0-2}) is given by
\begin{eqnarray}
E_g=E_g^0+\frac{\ln3}{2}t-2t+B_{|c|}t+B_{|d|}t.\label{E0-4}
\end{eqnarray}
The surface energy is
\begin{eqnarray}
E_{\mbox{surf}}&=&E_g-E_g^0=\frac{\ln3}{2}t+\frac{1}{3}[{\rm sgn}(c)+{\rm sgn}(d)-4]t\no\\
&&-{\rm sgn}(c)B_{|c|}t-{\rm sgn}(d)B_{|d|}t.
\end{eqnarray}
\subsection{The case of $t+\xi_N=|\vec h_N|$ and $t+\xi_1=-|\vec h_1|$}
In this case, the Bethe ansatz equation (\ref{BA-4-1}) will become
\begin{eqnarray}
&&(\frac{\lambda_j-\frac{\eta}{2}}{\lambda_j+\frac{\eta}{2}})^{2N}=-\frac{\lambda_j-g\eta}{\lambda_j+g\eta}\prod_{l=1}^m\frac{\lambda_j-\theta_l-\frac{\eta}{2}}{\lambda_j-\theta_l+\frac{\eta}{2}}
\frac{\lambda_j+\theta_l-\frac{\eta}{2}}{\lambda_j+\theta_l+\frac{\eta}{2}},\no\\
&&\quad\quad\quad j=1,2,\ldots,M,\label{BA-4-9}
\end{eqnarray}
with
\begin{eqnarray}
g=\frac{\xi_N}{2|\vec{h}_N|}.\label{f}
\end{eqnarray}
We find that now there does exist a boundary string located at $\lambda_0=g\eta$ in the thermodynamics limit when $\xi_N>0$. Suppose that the density of $\lambda$ and $\theta$ in the state with boundary string are $\bar{\rho}(\lambda)$ and $\bar{\sigma}(\theta)$ and the density of $\lambda$ and $\theta$ in the state without boundary string are ${\rho}(\lambda)$ and ${\sigma}(\theta)$, thus we have
\begin{eqnarray}
a_1(\lambda)-\frac{1}{2N}\delta(\lambda)&=&\int_{-\infty}^{\infty}a_1(\lambda-x)\bar{\sigma}(x)dx+\bar{\rho}(\lambda)
+\frac{1}{2N}a_{2g}(\lambda),\label{BA-4-10}
\end{eqnarray}
and
\begin{eqnarray}
&&\frac{1}{2N}[a_1(\theta)+a_{2c}(\theta)+a_{2d}({\theta})]+\int_{-\infty}^{\infty}a_1(\theta-y)\bar{\rho}(y)dy
+\frac{1}{2N}[a_1(\theta+\lambda_0)+a_1(\theta-\lambda_0)]\no\\
&&\quad\quad=\frac{1}{2N}\delta(\theta)+\bar{\sigma}(\theta)
+\int_{-\infty}^{\infty}a_2(\theta-y)\bar{\sigma}(\theta)dy.\label{BA-4-11}
\end{eqnarray}
Then we have
\begin{eqnarray}
\int_{-\infty}^{\infty}a_1(\lambda-x)\delta\sigma(x)dx+\delta\rho(\lambda)=0,\label{BA-4-12}
\end{eqnarray}
\begin{eqnarray}
&&\int_{-\infty}^{\infty}a_2(\theta-y)\delta\sigma(y)dy+\delta\sigma(\theta)-\hspace{-0.1truecm}\int_{-\infty}^{\infty}a_1(\theta-y)\delta\rho(y)dy\no\\
&&\quad\quad=\frac{1}{2N}[a_1(\theta+\lambda_0)+a_1(\theta-\lambda_0)],\label{BA-4-13}
\end{eqnarray}
with
\begin{eqnarray}
\delta\rho(\lambda)=\bar{\rho}(\lambda)-\rho(\lambda),\quad\quad\delta\sigma(\theta)=\bar{\sigma}(\theta)-\sigma(\theta).\no
\end{eqnarray}
Using the Fourier expansion, we have
\begin{eqnarray}
&&\quad\hspace{-1.0truecm}\tilde{a}_1(\omega)\delta\tilde{\sigma}(\omega)+\delta\tilde{\rho}(\omega)=0,\label{BA-4-14}\\
&&\hspace{-1.0truecm}[\tilde{a}_2(\omega)+1]\delta\tilde{\sigma}(\omega)-\tilde{a}_1(\omega)\delta\tilde{\rho}(\omega)=\frac{1}{2N}A(\omega),\label{BA-4-15}
\end{eqnarray}
with $A(\omega)=\tilde{a}_{2g+1}(\omega)-\tilde{a}_{2g-1}(\omega)$ if $g>\frac{1}{2}$ and $A(\omega)=\tilde{a}_{1-2g}(\omega)+\tilde{a}_{2g+1}(\omega)$ if $0<g<\frac{1}{2}$.
Solving equations (\ref{BA-4-14}) and (\ref{BA-4-15}) gives rise to that
\begin{eqnarray}
\delta\tilde{\rho}(\omega)=-\frac{1}{2N}\frac{\tilde{a}_1(\omega)A(\omega)}{2\tilde{a}_2(\omega)+1}.
\end{eqnarray}
When $g>\frac{1}{2}$, which is corresponding to $t<0$, the difference between the energy of the state with the string and $E_g$ is
\begin{eqnarray}
\Delta E_1&&\hspace{-0.1truecm}=\hspace{-0.1truecm}-2Nt\int_{-\infty}^{\infty}\delta\tilde{\rho}(\omega)[\delta(\omega)-\frac{1}{2}\tilde{a}_1(\omega)]d\omega-2t\frac{\lambda_0^2-\frac{1}{4}}{\lambda_0^2+\frac{1}{4}}\no\\
&&\hspace{-0.1truecm}=\hspace{-0.1truecm}-2t-\frac{t}{g^2-\frac{1}{4}}+tB_{g-{\frac{1}{2}}}-tB_{g+\frac{1}{2}}>0.\label{EO-6}
\end{eqnarray}
When $0<g<\frac{1}{2}$, which is corresponding to $t>0$, the difference between the energy of the state with the string and $E_g$ is
\begin{eqnarray}
\Delta E_2&&\hspace{-0.1truecm}=\hspace{-0.1truecm}-2Nt\int_{-\infty}^{\infty}\delta\tilde{\rho}(\omega)[\delta(\omega)-\frac{1}{2}\tilde{a}_1(\omega)]d\omega-2t\frac{\lambda_0^2-\frac{1}{4}}{\lambda_0^2+\frac{1}{4}}\no\\
&&\hspace{-0.1truecm}=\hspace{-0.1truecm}-\frac{4}{3}t+\frac{t}{\frac{1}{4}-g^2}-tB_{{\frac{1}{2}}-g}-tB_{\frac{1}{2}+g}>0.\label{EO-7}
\end{eqnarray}
We can obtain the similar result when $t+\xi_N=-|\vec h_N|$ and $t+\xi_1=|\vec h_1|$, which shows that the correct ground state contains only real roots when the two boundary fields $\vec{h}_1$ and $\vec{h}_N$ are parallel. The surface energy in this case is given by
\begin{eqnarray}
E_{\mbox{surf}}&=&\frac{\ln3}{2}t+\frac{1}{3}[2{\rm sgn(g)}+{\rm sgn}(c)+{\rm sgn}(d)-4]t\no\\
&&-[{\rm sgn}(c)B_{|c|}-{\rm sgn}(d)B_{|d|}+{\rm sgn}(g)B_{|g|-\frac{1}{2}}\no\\
&&+{\rm sgn}(g)B_{|g|+\frac{1}{2}}]t.\label{E0-5}
\end{eqnarray}\label{E0-51}

\section{Conclusion}

The one-dimensional super-symmetric t-J model with unparallel
boundary magnetic fields described by the Hamiltonian (\ref{Ham}) is
studied by combining the coordinate Bethe ansatz and off-diagonal
Bethe ansatz methods. With the coordinate Bethe ansatz,
eigen-functions of the Hamiltonian of the model are given in terms
of some quasi-momentum $\{k_j\}$ as (\ref{Wave-function}). The
constraints (\ref{bar-tau}) on these quasi-momentum is transformed
into the eigenvalues problem of the resulting transfer matrix of the
associated open XXX spin chain with arbitrary boundary fields. The
second eigenvalue problem is then solved via the off-diagonal Bethe
ansatz method.   We remark that further study on the correlation
functions would be an interesting issue.

\section*{Acknowledgments}

The financial support from  the National Natural Science Foundation
of China (Grant Nos. 11174335, 11031005, 11375141,
11374334), the National Program for Basic Research of MOST (973
project under grant No.2011CB921700), the State Education
Ministry of China (Grant No. 20116101110017) and BCMIIS are
gratefully acknowledged.

\end{document}